\documentclass[aps,prb,floatfix,amsmath,amssymb,preprint,eqsecnum,nofootinbib]{revtex4}
\usepackage{graphicx}% Include figure files
\usepackage{dcolumn}% Align table columns on decimal point
\usepackage{bbm}% bold math
\usepackage{epstopdf}
\usepackage{color}

\usepackage{setspace} %line spacing
\newcommand{\beq}{\begin{equation}}
\newcommand{\eeq}{\end{equation}}
\newcommand{\beqn}{\begin{eqnarray}}
\newcommand{\eeqn}{\end{eqnarray}}

\usepackage{amsmath}
\usepackage{amssymb}

\def \q{{\mathbf{q}}}
\def \R{{\mathbf{R}}}

\def \k{{\mathbf{k}}}
\def \K{{\mathbf{K}}}
\def \q{{\mathbf{q}}}
\def \p{{\mathbf{p}}}

\def \R{{\mathbf{R}}}

\def \S{{\mathbf{S}}}
\def \s{{\mathbf{s}}}

\begin{document}
%\preprint{arXiv:1105.xxxx}

\title{Fermi surface reconstruction in hole-doped $t$-$J$ models\\ without long-range antiferromagnetic order.}

\author{Matthias Punk}
\affiliation{Department of Physics, Harvard University, Cambridge MA
02138}

\author{Subir Sachdev}
\affiliation{Department of Physics, Harvard University, Cambridge MA
02138}

\date{\today \\
\vspace{1.6in}}

\begin{abstract}
We calculate the Fermi surface of electrons in hole-doped, extended $t$-$J$ models on a square lattice in a regime where no long-range antiferromagnetic order is present, and no symmetries are broken. Using the ``spinon-dopon'' formalism of Ribeiro and Wen,
we show that short-range antiferromagnetic correlations lead to a reconstruction of the Fermi surface into hole pockets which are  not necessarily centered at the antiferromagnetic Brillouin zone boundary.
The Brillouin zone area enclosed by the Fermi surface is proportional to the density of dopants away from half-filling,
in contrast to the conventional Luttinger theorem which counts the total electron density.
This state realizes a ``fractionalized Fermi liquid'' (FL*), which has been proposed as a possible ground-state of the underdoped cuprates;
we note connections to recent experiments.
We also discuss the quantum phase transition from the FL* state to the Fermi liquid state with long-range antiferromagnetic order.

\end{abstract}

\maketitle

\section{Introduction}
\label{sec:intro}

The nature of electronic Fermi surfaces in strongly correlated metals, in particular underdoped cuprates, has been the subject of intensive debate for many years. Recent observations of pocket-like Fermi surfaces in quantum oscillation experiments\cite{Leyraud2007, Taillefer2009, Sebastian2010,ramshaw,Choi2011} as well as new angle resolved photo emission measurements\cite{Yang2011} have triggered a renewed theoretical interest in this matter\cite{Millis2007, Chakravarty2008, Zhang2008}.

One possible, well known route to a Fermi surface reconstruction is the onset of spin-density wave (SDW) order, which breaks a large Fermi surface into small electron- and hole-pockets centered at the magnetic Brillouin zone boundary\cite{Subir1995, Chubukov1997}.
In fact, many of the unresolved theoretical problems in strongly correlated electron materials, from heavy-Fermion compounds to high-$T_c$ cuprates, are related to the fate of electronic excitations close to antiferromagnetic quantum critical points\cite{Max2010}.
It has been been argued, however, that the critical point between a metal with a large Fermi surface and an antiferromagnetic metal with small Fermi pockets may be replaced by a new intermediate phase, the so called fractionalized Fermi liquid (FL*) \cite{Senthil1, Senthil2}, which exhibits small pockets similar to the antiferromagnetic metal, but breaks no symmetries: summaries of these arguments, and of previous theoretical work, can be found in two recent reviews.\cite{vojtarev,sces}

The simplest picture of the FL* phase appears in the context of Kondo lattice models coupling a lattice of localized $f$ moments
and a conduction band of itinerant $c$ electrons. There are two important energy scales to consider: the Kondo exchange $J_K$ between
the $f$ moments and the $c$ electrons, and Heisenberg exchange $J_H$ between the $f$ moments. If $J_K \gg J_H$, then $f$ moments are ``Kondo-screened'' by the conduction electrons, leading to a Fermi liquid ground state with Fermi surfaces enclosing the traditional Luttinger volume which counts the density of {\em both\/} the $f$ and $c$ electrons; the only memory of the localized nature
of the underlying $f$ electrons is the that electronic quasiparticles near the Fermi surface have an effective mass which is much larger than the bare electron mass, and so this phase is often referred to as a `heavy' Fermi liquid. However, in the opposite parameter regime
$J_H \gg J_K$ other phases can appear. The most natural possibility is the appearance of magnetic order of the $f$ moments, but let us assume the $f$-$f$ couplings are sufficiently frustrated so that this does not happen. Then the $f$ moments may form a spin liquid, which does not break any symmetries of the lattice Hamiltonian. The formation of the spin liquid also quenches the Kondo effect, and so the
effective value of $J_K$ does {\em not\/} renormalize to infinity as it does in the single impurity Kondo model \cite{bgg,Senthil1,Senthil2}.
The $c$ electrons are now only weakly coupled to the $f$ spin liquid, and so the $c$ electrons 
form a ``small'' Fermi surface which encloses a volume
controlled only by the density of $c$ electrons, which violates the traditional Luttinger count. This is the FL* metal.

This paper describes a FL* state in a {\em single\/}-band model appropriate for the cuprate superconductors. Previous studies\cite{Qi,EunGook} realized such a state by initially fractionalizing the electron into a neutral $S=1/2$ spinon, and a spinless
``holon'' carrying electromagnetic charge. The spinons eventually became the excitations of a `background' spin liquid,
analogous the spin liquid of the $f$ electrons above. And the holons eventually captured a spinon to reconstitute as
electron-like particles which occupied the states inside a small Fermi surface. Because of this somewhat intricate sequence of
transformations, the description of the FL* state was only achieved in a semi-phenomenological manner. 

Here we will provide a more direct and quantitative description of the FL* state in a single-band model. The key step will be a rewriting
of the single band degrees of freedom in a manner which mimics those of the Kondo lattice. Such a formulation is provided by the representation of Ribeiro and Wen\cite{Ribeiro} in which the electron fractionalizes into a neutral spinon and a ``dopon'' which has the same
quantum numbers as the electron.

The rest of the paper is organized as follows. In section \ref{sec:models} we introduce the extended $t$-$J$ model in the representation of Ribeiro and Wen\cite{Ribeiro}, which is ideally suited for our purposes. Section \ref{sec:flstar} deals with our approach to construct FL* ground states and presents results for the shape and position of the electronic Fermi surface. 
Section~\ref{sec:transition} describes the Fermi surface evolution from the FL* state to the Fermi liquid state with long-range
antiferromagnetic order, along with a discussion of the quantum-critical properties.
We summarize our results, and note connections to recent experiments in Section~\ref{sec:conc}.

\section{Model}
\label{sec:models}

In the following we want to study ground-states of extended $t$-$J$ Hamiltonians on the square lattice
\begin{equation}
H = - \frac{1}{2} \sum_{ij} t_{ij} \left( \tilde{c}^\dagger_{i\sigma} \tilde{c}^{\ }_{j\sigma} + \text{h.c} \right) + \frac{1}{2} \sum_{ij} J_{ij} \left( \s_i \cdot \s_j - \frac{1}{4} n_i n_j \right) \ ,
\label{HtJ}
\end{equation}
where $\tilde{c}^\dagger_{i\sigma}$  ($ \tilde{c}^{\ }_{i\sigma}$) denotes the Gutzwiller projected creation (annihilation) operator of electrons with spin $\sigma$ on lattice site $i$, $\s_i = \tilde{c}^\dagger_{i\alpha} \boldsymbol{\sigma}_{\alpha \beta} \tilde{c}^{\ }_{i\beta}$  is the electron spin operator and $n_i = \tilde{c}^{\dagger}_{i\sigma} \tilde{c}^{\ }_{i\sigma}$ the electron number operator (here and in the following we sum over repeated spin indices). 
We are interested in describing possible ground states slightly below half filling $n=1-x$, where the density of doped holes is small $x \ll 1$ but large enough to destroy any long-range magnetic order. In addition these ground states should not break any lattice symmetries. In particular we want to show that strong short-range antiferromagnetic correlations already lead to a reconstructed Fermi surface consisting of small hole pockets, the area of which is proportional to the dopant density $x$, instead of $1-x$ as for conventional Fermi liquids. Such ground states realize a fractionalized Fermi liquid\cite{Senthil1, Senthil2}.
 
Our starting point is the spinon-dopon formulation of the $t$-$J$ model developed by Ribeiro and Wen \cite{Ribeiro}. In this representation the elementary excitations are spinons, which carry spin-1/2 but no charge, and dopons, which carry spin-1/2 and charge. Accordingly, Ribeiro and Wen introduce two degrees of freedom per lattice site, a 'localized' spin-1/2  as well as a fermionic spin-1/2 degree of freedom - the dopon - representing a doped charge carrier. A physical hole corresponds to  a singlet of a lattice spin and a dopon. The correspondence between single-site basis states is shown in Tab.~\ref{corresp}. Following this approach, the $t$-$J$ Hamiltonian in Equ.~\eqref{HtJ} takes the form\cite{Ribeiro}
\begin{eqnarray}
H &=& \frac{1}{2}  \sum_{ij} J_{ij} (\mathbf{S}_i \cdot \mathbf{S}_j-1/4) \,  \mathcal{P} (1-d^\dagger_{i \alpha} d^{\ }_{i \alpha}) (1-d^\dagger_{j \beta} d^{\ }_{j \beta}) \mathcal{P} \notag \\
&& + \frac{1}{2}\sum_{ij} \frac{t_{ij}}{2} \mathcal{P} \Big[ \frac{1}{4} d_{i \alpha}^\dagger d_{j \alpha}^{\ }   -\frac{1}{2} (d^\dagger_{i \alpha} \vec{\sigma}_{\alpha \beta} d_{j \beta}) \cdot (\mathbf{S}_i+\mathbf{S}_j) +d^\dagger_{i \alpha} d^{\ }_{j \alpha} \mathbf{S}_i \cdot \mathbf{S}_j  \notag \\
&& + \,  i \, (d^\dagger_{i \alpha} \vec{\sigma}_{\alpha \beta} d^{\ }_{j \beta}) \cdot (\mathbf{S}_i \times \mathbf{S}_j ) +\text{h.c.} \Big] \mathcal{P} -\mu \sum_i  d^\dagger_{i \alpha} d^{\ }_{i \alpha} \ .
\label{h0}
\end{eqnarray}
Here $\mathcal{P} = \prod_j (1-d^\dagger_{j \uparrow} d^{\ }_{j \uparrow} d^\dagger_{j \downarrow} d^{\ }_{j \downarrow} )$ denotes the Gutzwiller projector for the fermionic spin-1/2 operators $d^\dagger_i$ and $d_i$ that create or annihilate a dopon on lattice site $i$, and we added a chemical potential $\mu$ for the dopons. Note again that the spins $\mathbf{S}_j$ on each lattice site $j$ are independent, localized spin-1/2 degrees of freedom and are not associated with the spin of the dopons. This representation of the $t$-$J$ model is faithful in the sense that the Hamiltonian doesn't couple the physical singlet- and the unphysical triplet states in the enlarged Hilbert space that is spanned by the spin- and the dopon degree of freedom \cite{Ribeiro}. A projection to the physical Hilbert space is thus not necessary.

\begin{table}
\caption{Single site basis-state correspondence: t-J \emph{v.s.} spinon-dopon.}
\begin{center}
\begin{tabular}{c|c}
\text{t-J} & \text{spinon-dopon} \\
\hline \\
$| \! \uparrow \rangle_i$ &  $| \! \uparrow 0 \rangle_i$ \\
$| \! \downarrow \rangle_i$ &  $| \! \downarrow 0 \rangle_i$ \\
$| 0 \rangle_i$ &  $( | \! \uparrow \downarrow \rangle_i - | \! \downarrow \uparrow \rangle_i)/\sqrt{2} $ \\
unphys. \ & \ triplet-states \\
unphys. \ &  \ doubly occupied dopon
\end{tabular}
\end{center}
\label{corresp}
\end{table}

In terms of the spin- and dopon operators the Gutzwiller projected electron operators take the form
\begin{equation}
\tilde{c}^\dagger_{j \sigma} = \frac{\sigma}{\sqrt{2}} \mathcal{P}  \left(  (1/2 + \sigma S^z_j) d_{j -\sigma} - S^{\sigma}_j d_{j \sigma} \right) \mathcal{P} \ ,
\label{ctilde}
\end{equation}
where $S^\sigma$ denotes the spin raising (lowering) operator $S^+$ ($S^-$) for $\sigma=\uparrow \, (\downarrow)$.
From Equ.~\eqref{ctilde} one can easily show that total density of electrons is given by
\begin{equation}
\sum_\sigma  \tilde{c}^\dagger_{j \sigma} \tilde{c}^{\ }_{j \sigma} =  \mathcal{P} ( 1 - \sum_\sigma d^\dagger_{j \sigma} d^{\ }_{j \sigma} ) \mathcal{P} \overset{\text{low doping}}{\approx} 1 - \sum_\sigma d^\dagger_{j \sigma} d^{\ }_{j \sigma} \ ,
\label{Cdensity}
\end{equation}
i.e.~the density of doped charge carriers equals the density of dopons $x=\sum_\sigma \langle d^\dagger_{i \sigma} d_{i \sigma} \rangle$, as expected. 

As will be explained in more detail below, our main assumption  is that the localized lattice spins form a $\mathbb{Z}_2$ spin liquid with bosonic spinon excitations. This is reasonably justified in the doping regime close to the antiferromagnetically ordered phase, where the interaction between the lattice spins is frustrated by the motion of dopons. The bosonic nature of the spinons prohibits a hybridization of spinons with fermionic dopons and gives rise to an electronic Fermi surface, the volume of which is determined by the density dopons $x$ alone, as long as no pairing instabilities occur. This is in contrast to the conventional Luttinger theorem, which states that in a metal without broken symmetries the 'volume' enclosed by the Fermi surface should be proportional to the total density of electrons $1-x$. It has been shown earlier, however, that topological excitations associated with the emergent gauge field of a spin liquid have to be included in the Luttinger count\cite{Senthil2}, giving rise to a FL* phase with small pocket Fermi surfaces the total volume of which is the same as in an antiferromagnetic metal. In the present formalism, this modified Luttinger theorem of a Fermi surface of size $x$ can be easily
proved by applying the usual many-body formalism to the system of interacting spinons and dopons described by Eq.~(\ref{h0})
(and more explicitly in Eq.~(\ref{hss}) below); we need only assume that the final state is adiabatically connected to a state of weakly
interacting spinons and dopons, and then the standard proof leads here to the novel Luttinger count of $x$.

In the presence of strong local AF correlations the most important couplings between the dopons and the localized spins are the two interaction terms in the second line of Equ.~\eqref{h0}. The $\sim \S_i \cdot \S_j$ term leads to a strong suppression of nearest neighbor hopping of dopons in a locally AF ordered background, whereas the $\sim (\S_i + \S_j)$ term is responsible for scattering of dopons with momentum transfer close to $\q=(\pi,\pi)$. 
In the following we are going to neglect the $\sim \S_i \times \S_j$ term due to the expected strong local collinear AF correlations. Also, we use a mean-field decoupling of the Heisenberg exchange term in the first line of \eqref{h0}, i.e. $J \to (1-x)^2 J$ and drop the Gutzwiller projectors, which is safe in the low doping limit $x \ll 1$.

\section{FL* and electron Fermi surface}
\label{sec:flstar}

As mentioned above, the main prerequisite in order to get a fractionalized Fermi liquid is that the localized spins form a spin liquid.
Within our model \eqref{h0} spin liquid ground states can be conveniently described using a Schwinger-Boson representation for the lattice spins, i.e.~we write
\begin{equation}
\mathbf{S}_i = \frac{1}{2} b^\dagger_{i \alpha} \boldsymbol{\sigma}_{\alpha \beta} b^{\ }_{i \beta} \ ,
\label{sb}
\end{equation}
which requires the constraint $\sum_\sigma b^\dagger_{i \sigma} b_{i \sigma} = 1$ to hold on every lattice site. Note that there is an emergent $U(1)$ gauge structure associated with the redundancy of the Schwinger Boson representation under local phase transformations $b_{j \sigma} \to b_{j \sigma} \exp(i \phi_j)$. The exchange terms can be expressed in terms of Schwinger Bosons using the identity
\begin{equation}
\mathbf{S}_i \cdot \mathbf{S}_j = -\frac{1}{2} (\epsilon_{\alpha \beta} b^\dagger_{i \alpha} b^{\dagger }_{j \beta}) (\epsilon_{\gamma \delta} b^{\ }_{i \gamma} b^{\ }_{j \delta}) +\frac{1}{4}+\frac{\delta_{ij}}{2} \ .
\label{h01}
\end{equation}
Inserting these expressions in the Hamiltonian \eqref{h0} and using the approximations mentioned above, we obtain the Hamiltonian
\begin{eqnarray}
H &=& -\frac{1}{4} \sum_{ij} \left[ J_{ij} + \frac{t_{ij}}{2} (d^\dagger_{i \alpha} d^{\ }_{j \alpha}+d^\dagger_{j \alpha} d^{\ }_{i \alpha}) \right]  \, \epsilon_{\alpha \beta} b^\dagger_{i \alpha} b^{\dagger }_{j \beta}  \, \epsilon_{\gamma \delta} b^{\ }_{i \gamma} b^{\ }_{j \delta}  \label{hss} \\
&&+\frac{1}{2} \sum_{ij} \frac{t_{ij}}{4} \left[ 2 \, d^\dagger_{i \alpha} d^{\ }_{j \alpha} - d^\dagger_{i \alpha} d^{\ }_{j \beta} \big( b^\dagger_{i \beta} b^{\ }_{i \alpha} + b^\dagger_{j \beta} b^{\ }_{j \alpha} \big) + \text{h.c.} \right] + \sum_i  \big( \lambda \, b^\dagger_{i \alpha} b^{\ }_{i \alpha} - \mu \, d^\dagger_{i \alpha} d^{\ }_{i \alpha} \big) \ , \notag
\end{eqnarray}
where $\lambda$ is the Lagrange multiplier that enforces the Schwinger Boson constraint on average. In this representation a spin liquid can be conveniently described by employing the mean-field decoupling
\begin{equation}
Q_{ij} = \frac{1}{2} \langle  \epsilon_{\alpha \beta} b^\dagger_{i \alpha} b^{\dagger }_{j \beta}  \rangle \ .
\end{equation}
By construction this mean-field decoupling preserves the $SU(2)$ invariance since $Q_{ij}$ is a singlet expectation value.
After a Fourier transformation we obtain the euclidean mean-field action (we use the shorthand notation $k = (\omega, \k)$)
\begin{eqnarray}
S_\text{MF}/\beta &=& \sum_{k,\sigma}  \bar{d}_{k \sigma} (-i \omega_n +\xi^0_\k) d_{k \sigma} +  \sum_{k} B^\dagger_{k} 
\begin{bmatrix}
-i \Omega_n + \lambda & - \sum_\p Q_\p J_{\p-\k} \\
- \sum_\p Q^*_\p J_{\p-\k}  & i \Omega_n + \lambda
\end{bmatrix} B_{k} \notag \\
&&- \sum_{q,k,k'} \bar{d}^{\ }_{k'+q-k \sigma} \, B^\dagger_{k} \, \mathbf{V}^{\sigma \sigma'}_{\k' \k \q} \, B^{\ }_{q} \, d^{\ }_{k' \sigma'} +  \sum_{\k \q} Q^*_{\k+\q} J_\q Q_\k   \notag \\
&& - \frac{1}{2} \sum_{q,k,k'} (t_{\k'}+t_{\k'+\q+\k}) \left[ B_{k \downarrow} B_{q \uparrow} \bar{d}_{k+k'+q \uparrow} d_{k' \downarrow} +  B^*_{k \uparrow} B^*_{q \downarrow} \bar{d}_{k' \downarrow} d_{k'+k+q \uparrow} \right] \ ,
\label{hmf}
\end{eqnarray}
where we have introduced the bosonic Nambu spinor 
\begin{equation}
B_{k} =\begin{pmatrix}
b_{k \uparrow} \\
b^*_{-k \downarrow}
\end{pmatrix} 
\end{equation}
and 
\begin{equation}
\mathbf{V}^{\sigma \sigma'}_{\k' \k \q} = \begin{bmatrix}
\frac{1}{2} (t_{\k'}+t_{\k'+\q-\k}) \delta_{\sigma,\uparrow} & \sum_p Q_\p (t_{\p+\k'-\k}+t_{\p-\k'-\q}) \\
\sum_p Q^*_\p (t_{\p+\k'-\k}+t_{\p-\k'-\q}) & \frac{1}{2} (t_{\k'}+t_{\k'+\q-\k}) \delta_{\sigma,\downarrow} 
\end{bmatrix} \delta_{\sigma \sigma'}  \ ,
\end{equation}
as well as
\begin{equation}
\xi_\k^0 =t_\k + 2 \sum_{\q \p} Q^*_\q t_\p Q_{\k+\q-\p} - \mu \ ,
\label{xi0}
\end{equation}
where $t_\k$ is the usual $t t' t''$ tight-binding dispersion on the square lattice with non-zero hopping amplitudes up to third nearest neighbors
\begin{equation}
t_\k =  t \, \big( \cos k_x + \cos k_y \big) + t' \, 2 \cos k_x  \cos k_y +  t'' \, \big( \cos 2 k_x + \cos 2 k_y \big) \ .
\end{equation}
Note that in contrast to $J_\k$ and $Q_\k$ we have absorbed a factor $1/2$ in the definition of $t_\k$. Moreover, $i \omega_n$ and $i \Omega_n$ denote fermionic and bosonic Matsubara frequencies, respectively. 
Because the bosonic spinon modes are gapped in the spin-liquid phase we can safely integrate them out and obtain an effective action for the dopon fields $d$. Expanding to second order in the bosonic propagator we get
\begin{eqnarray}
S^{(2)}_\text{MF}[\bar{d}, d] &=& S_0[\bar{d}, d] + \text{Tr} \log \beta \mathcal{G}^{-1}_0 - \text{Tr} \, \mathcal{G}_0 \Phi \notag \\
&& -\frac{1}{2} \text{Tr} \, \mathcal{G}_0 \Phi  \mathcal{G}_0 \Phi - \text{Tr} \, \mathcal{G}_0 D  \mathcal{G}_0 \bar{D} -\text{Tr} \, \mathcal{G}_0 \bar{D}  \mathcal{G}_0 D \ ,
\label{s2}
\end{eqnarray}
where $\text{Tr}$ denotes the trace with respect to momentum-, Matsubara- and Nambu indices. Furthermore we have defined 
\begin{equation}
(\mathcal{G}^{-1}_0)_{kq} = \begin{bmatrix}
-i \Omega_n + \lambda & - \sum_\p Q_\p J_{\p-\k} \\
- \sum_\p Q^*_\p J_{\p-\k}  & i \Omega_n + \lambda
\end{bmatrix} \delta_{kq}
\end{equation}
and
\begin{eqnarray}
\Phi_{kq} &=& \sum_{k'} \bar{d}_{k'+q-k \, \sigma} \mathbf{V}^{\sigma \sigma'}_{\k' \k \q}  d_{k' \sigma'}  \\
D_{kq} &=&  \frac{1}{4} \sum_{k'} (t_{\k'}+t_{\k'+\q+\k}) \bar{d}_{k'\downarrow} d_{k'+k+q \uparrow} \ \sigma_x \\
\bar{D}_{kq} &=&  \frac{1}{4} \sum_{k'} (t_{\k'}+t_{\k'+\q+\k}) \bar{d}_{k'+k+q\uparrow} d_{k'\downarrow} \ \sigma_x \ ,
\end{eqnarray}
where $\sigma_x$ denotes the respective Pauli matrix in Nambu space. 
The effective dopon action in Equ.~\eqref{s2} describes the hopping of doped charge carriers in a locally AF-ordered background as well as the residual interactions between dopons due to the exchange of a spinon pair. Note, however, that the Schwinger boson mean-field theory presented here cannot be used to describe a conventional Fermi liquid state with a large Fermi surface at large doping, where the Luttinger volume is determined by the total density of electrons $1-x$. This phase can be described using Schwinger Fermions instead of Bosons \cite{Ribeiro}. In this case a hybridization between dopons and Fermionic spinons can lead to a ground-state with a large Fermi surface.

\subsection{Gaussian dopon action}

The linear contribution $\sim \mathcal{G}_0$ to the effective dopon action \eqref{s2} basically descends from the term in the Hamiltonian \eqref{h0} which couples the dopons to $\S_i \cdot \S_j$ and thus strongly renormalizes the bare dopon dispersion. Performing the Matsubara summation and the trace over the Nambu indices of the $\text{Tr} \, \mathcal{G}_0 \Phi $ term we get the Gaussian action
\begin{eqnarray}
S^{(1)}_\text{MF} &=& 
 \beta \sum_{k,\sigma}  \bar{d}_{k \sigma} \Big[ -i \omega_n +\xi^0_\k -\sum_{\k'} \frac{\lambda t_\k + 4 \sum_{\q \p} Q^*_{\p} Q_{\q} J_{\q-\k'} t_{\p+\k-\k'}}{2 E_{\k'}} \Big]  d_{k \sigma} \notag \\
 && + \text{Tr} \log \beta \mathcal{G}_0^{-1} + \beta \sum_{\k \q} J_{\q-\k} Q^*_\q Q_\k  \ .
 \label{s1}
\end{eqnarray}
This expression can be simplified using the self-consistency conditions for the Lagrange multiplier $\lambda$ as well as for the mean-field $Q_\k$. We make one further approximation here, however, and determine $\lambda$ and $Q_\k$ at the Gaussian level not fully self-consistent, but only with respect to $S^{(0)}_\text{MF} = \text{Tr} \log \beta \mathcal{G}^{-1}_0$, i.e.~we neglect the back-action of the dopons on the spinons. The approximate self-consistency equations thus read
\begin{eqnarray}
1 &=& \frac{\partial F^{(0)}}{\partial \lambda} \ = \ \sum_\k \frac{\lambda}{E_\k} \\
0 &=& \frac{\partial F^{(0)}}{\partial Q^*_\p} \ = \ \sum_\k J_{\p-\k} \left[ Q_\k - \frac{\sum_\q J_{\q-\k} Q_\q}{2 E_\k}\right]   \ ,
\end{eqnarray}
where $F^{(0)}$ is the free energy associated with $S^{(0)}_\text{MF}$. Here and in the following $E_\k$ denotes the spinon dispersion relation, which is given by
\begin{equation}
E_\k = \sqrt{\lambda^2 -|\sum_\q Q_\q J_{\q-\k}|^2} \ .
\label{disp0}
\end{equation}
Inserting these expressions back into \eqref{s1} we get
\begin{equation}
S^{(1)}_\text{MF} = \beta \sum_{k,\sigma}  \bar{d}_{k \sigma} \big[ -i \omega_n + \xi_\k \, \big]  d_{k \sigma} + \text{Tr} \log \beta \mathcal{G}_0^{-1} + \text{const.}
\end{equation}
with the Gaussian dopon dispersion
\begin{equation}
\xi_\k =t_\k/2 - 2 \sum_{\q \p} Q_\q Q^*_{\p} t_{\p+\k-\q} - \mu \ .
\label{xikrenorm}
\end{equation}
Note the different sign of the second term compared to the bare dopon dispersion $\xi_\k^0$ in Equ.~\eqref{xi0}.

\subsection{Mean-field ansatz for a $\mathbb{Z}_2$-FL*}

In the following we use the simplest mean field ansatz for the $Q_{ij}$'s. We take a zero-flux state with $Q_{ij}=Q$ on nearest neighbor bonds. The invariant gauge group\cite{Wen2002} (IGG), i.e.~the group of gauge transformations that leaves this ansatz invariant, is $U(1)$ due to the bipartite nature of the square lattice. Indeed, we can choose the gauge transformation $b_{j \sigma} \to b_{j \sigma} \exp( i \phi)$ on sublattice $A$ and $b_{j \sigma} \to b_{j \sigma} \exp(-i \phi)$ on sublattice $B$ without changing the ansatz. However, since gapless $U(1)$ gauge fluctuations are hard to control, we break the $U(1)$ gauge group down to $\mathbb{Z}_2$ by including frustration in the form of a small next-nearest neighbor exchange interaction $J'$ as well as a corresponding singlet bond-amplitude $Q'$. The excitations of the emergent $\mathbb{Z}_2$ gauge field are gapped visons which shouldn't play a big role in our subsequent analysis as long as their gap is sufficiently large, thus we neglect them in the following. The Fourier transformation of our ansatz $Q_{ij}$ then takes the form
\begin{equation}
Q_\k =  i 2 \left[ Q (\sin k_x + \sin k_y) + Q' ( \sin(k_x+k_y) + \sin(k_y-k_x)) \right] \ .
\label{ansatz}
\end{equation}
Note that $Q_{ji} = -Q_{ij}$ and thus $Q_{-\k}= -Q_{\k}$. The corresponding spinon dispersion relation \eqref{disp0} is given by
\begin{equation}
E_\k  = \sqrt{\lambda^2 - 4 \,  | J Q (\sin k_x + \sin k_y) + 2 J' Q'  \cos k_x \, \sin k_y|^2} \ ,
\label{Ek}
\end{equation}
where $J$ and $J'$ are the nearest- and next-nearest-neighbor exchange couplings whereas $Q$ and $Q'$ are the corresponding singlet amplitudes on nearest- and next-nearest neighbor bonds.
For this choice of $Q_\k$ the convolutions in all the expressions for the effective action can be evaluated straightforwardly. In particular, the Gaussian dopon dispersion from Eq.~\eqref{xikrenorm} takes the form
\begin{eqnarray}
\xi_\k &=& \frac{t}{2} (1-4 |Q|^2 )\, \big( \cos k_x + \cos k_y \big) \notag \\
&& + \, t' (1-4 |Q'|^2 ) \, \cos k_x  \cos k_y +  \frac{t''}{2} \, \big( \cos 2 k_x + \cos 2 k_y \big)-\mu \ .
\label{dopdisp}
\end{eqnarray}
Note that the dispersion is invariant under $\mathbb{Z}_2$ gauge transformations $Q_{ij} \to -Q_{ij}$. The singlet amplitudes $Q$ and $Q'$ can take values between $Q,Q' \in \big[0, \ 1/\sqrt{2}\big]$, where $Q=1/\sqrt{2}$ if nearest neighbor spins form a singlet. One can clearly see that the nearest neighbor hopping amplitude vanishes for perfect classical local AF correlations ($Q$=1/2) and it changes sign for $Q>1/2$. It is important to emphasize, however, that the Gaussian dopon dispersion \eqref{dopdisp} is strongly renormalized by the residual interaction.

\subsection{Residual interactions, effective dopon action at quartic order}

\begin{figure}
\begin{center}
\includegraphics[width= .7 \textwidth]{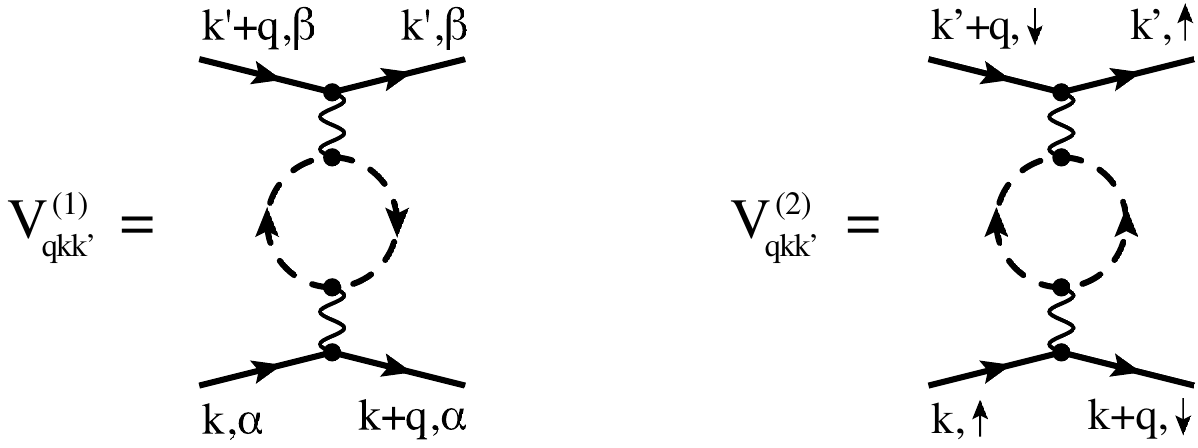}
\end{center}
\caption{Dopon-dopon interactions induced by the exchange of a spinon pair.}
\label{figInt}
\end{figure}

Here we analyze the interactions between dopons that are induced by the exchange of a spinon pair. The quadratic terms $\sim \mathcal{G}_0^2$ in the effective dopon action \eqref{s2} are given by
\begin{equation}
S_\text{int}^{(2)} = -\sum_{k q} \text{Tr}_2 \Big[ \frac{1}{2}(\mathcal{G}_{0})_k \Phi_{kq} (\mathcal{G}_{0})_q  \Phi_{qk} + (\mathcal{G}_{0})_k D_{kq} (\mathcal{G}_{0})_q  \bar{D}_{qk} + (\mathcal{G}_{0})_k \bar{D}_{kq} (\mathcal{G}_{0})_q  D_{qk} \Big]  \ .
\end{equation}
The first term gives rise to non-spinflip interactions, which we denote by $V^{(1)}_{\q\k\k'}$, whereas the second and third terms describe interactions where the dopon spins are flipped (denoted by $V^{(2)}_{\q\k\k'}$). Both interactions are shown schematically  in Fig.~\ref{figInt}.

\begin{figure}
\begin{center}
\includegraphics[width= \textwidth]{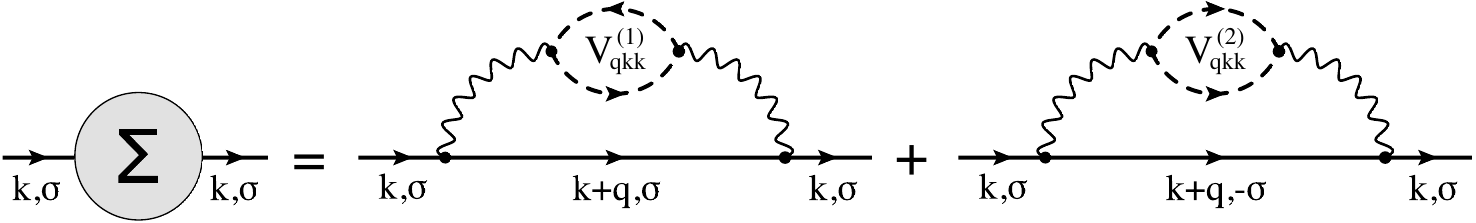}
\end{center}
\caption{Fock-type self-energy diagrams for the dopons.}
\label{figFock}
\end{figure}

In the following we perform a self-consistent Hartree-Fock analysis of this induced retarded interaction. Self-consistency is necessary because the interactions are strong and the shape as well as the position of the Fermi surface is strongly affected by interaction induced fluctuations\cite{Kane}. 
The Hartree-type interactions are accounted for already to a large extent in the Gaussian dopon dispersion \eqref{dopdisp}. In fact, the Hartree diagrams would correspond to a self-energy correction of the bosonic spinon propagator and are not expected to change the results qualitatively. For this reason we restrict ourselves to the two Fock-type diagrams shown in Fig.~\ref{figFock}. 
Since we expect the Fermi-liquid character of the dopons to prevail, we use a dominant pole approximation and neglect the incoherent part of the dopon Green's function. In this approximation, the dopon Green's function in the diagrams in Fig.~\ref{figFock} takes the form 
\begin{equation}
G_\sigma(\k,i \omega) \approx \frac{Z_\k}{-i \omega+\xi_\k} \ ,
\label{dopGF}
\end{equation}
where the quasiparicle residue $Z_\k$ as well as the dopon dispersion $\xi_\k$ are calculated self-consistently. We will justify this approximation \emph{a posteriori} by checking that the quasiparticle weight $Z_\k$ is reasonably large.
The dopon self-energy corresponding to the diagrams in Fig.~\ref{figFock} thus  takes the form
\begin{eqnarray}
\Sigma(\k,i \omega_n) &=& \frac{1}{\beta} \sum_q Z_{\k+\q} \, \frac{V^{(1)}_{qkk}+V^{(2)}_{qkk}}{-i \omega_n -i \Omega_q + \xi_{\k+\q}} \\
&=& - \frac{1}{8 \beta^2} \sum_{q k'} Z_{\k+\q} \,  \frac{i \Omega_{k'} ( i \Omega_{k'} + i \Omega_{q}) \, a_{\k \k' \q} + b_{\k \k' \q} }{[-i \omega_n -i \Omega_q + \xi_{\k+\q}] [ E_{k'}^2-(i \Omega_{k'})^2 ] [E_{k'+q}^2-(i \Omega_{k'}+i \Omega_{q})^2]} \notag \ ,
\end{eqnarray}
where $ a_{\k \k' \q}$ and  $b_{\k \k' \q}$ are momentum dependent factors given by
\begin{eqnarray}
a_{\k \k' \q} &=& 3 (t_\k+t_{\k+\q})^2 - 2 \big|(Q \! * \! t)_{\k'-\k}+(Q\! * \! t)_{\k'+\k+\q}\big|^2  \\
b_{\k \k' \q} &=& (t_\k+t_{\k+\q})^2 \big[3 \lambda^2+\big( (Q \! *\! J)_{\k'}  (Q \! * \! J)_{\k'+\q} + \text{c.c.} \big) \big] \notag \\
&&+\lambda (t_\k+t_{\k+\q})  \big[(Q^* \! * \! J)_{\k'}  + (Q^* \! *\! J)_{\k'+\q}\big] \big[(Q \! * \! t)_{\k'-\k}+(Q \! * \! t)_{\k'+\k+\q}\big] + \text{c.c.} \notag \\
&&+ (Q^* \! * \! J)_{\k'}  (Q^* \! * \! J)_{\k'+\q} \big[ (Q \! * \! t)_{\k'-\k}+(Q \! * \! t)_{\k'+\k+\q}\big] + \text{c.c.} \notag \\
&&+ 2 \lambda^2 \big|(Q \! * \! t)_{\k'-\k}+(Q \! * \! t)_{\k'+\k+\q}\big|^2    \ .
\end{eqnarray}
Here the asterisk denotes convolutions, i.e.~$(Q \! * \! t)_\k = \sum_\p Q_\p t_{\k-\p}$.
After performing the Matsubara sums and analytic continuation $i \omega_n \to \omega + i \delta$ we get (at $T=0$) 
\begin{eqnarray}
\Sigma_R(\k,\omega) &=& -\frac{1}{16} \sum_{\q, \k'} Z_{\k+\q} \, \Bigg\{  \frac{1}{2 E_{\k'} E_{\k'+\q}} \, \frac{E_{\k'} E_{\k'+\q} a_{\k \k' \q} - b_{\k \k' \q}}{\omega-E_{\k'}-E_{\k'+\q}-\xi_{\k+\q} + i \delta } \notag \\
&& \hspace{2.5cm} - \Theta(-\xi_{\k+\q}) \Bigg[ \frac{(E_{\k'}-\xi_{\k+\q}+\omega) a_{\k \k' \q}+ b_{\k \k' \q}/E_{\k'}}{E_{\k'+\q}^2-(E_{\k'}-\xi_{\k+\q}+\omega + i \delta)^2} \notag \\ 
&& \hspace{2.5cm} + \frac{(E_{\k'+\q}+\xi_{\k+\q}-\omega) a_{\k \k' \q} + b_{\k \k' \q}/E_{\k'+\q}}{E_{\k'}^2-(E_{\k'+\q} + \xi_{\k+\q} - \omega - i \delta)^2} \Bigg]   \Bigg\} \ .
\label{selfen}
\end{eqnarray}
The imaginary part of the retarded dopon self energy finally takes the form
\begin{eqnarray}
\text{Im} \Sigma_R(\k,\omega) &=& -\frac{\pi}{16} \sum_{\k' \q} Z_{\k+\q} \Big( \frac{b_{\k \k' \q}}{2 E_{\k'} E_{\k'+\q}} -\frac{a_{\k \k' \q}}{2} \Big) \big[  \Theta(\xi_{\k+\q}) \delta(\omega-E_{\k'}-E_{\k'+\q}-\xi_{\k+\q}) \notag \\
&& + \Theta(-\xi_{\k+\q}) \delta(\omega + E_{\k'} + E_{\k'+\q}-\xi_{\k+\q}) \big] \ .
\label{imsigma}
\end{eqnarray}
Note that $\text{Im} \Sigma_R(\omega) \equiv 0$ for $-2 \Delta < \omega < 2 \Delta$, where $\Delta$ denotes the spinon gap.

In all subsequent calculations we do not determine the Lagrange multiplier $\lambda$ (which fixes the Schwinger-Boson constraint) self-consistently, but use it to fix the value of the spinon gap $\Delta$. Moreover, we use the nearest- and next-nearest neighbor singlet amplitudes $Q$ and $Q'$ as free parameters.
Our calculational procedure works as follows:
first, we calculate $\text{Im} \Sigma_R(\k,\omega) $ numerically using the adaptive Monte-Carlo integration algorithm \textsc{Miser}\cite{Galassi}, which is based on a recursive stratified sampling method. 
Since the computational effort increases considerably with increasing accuracy, we set the bound of the relative error estimate to be smaller than 6\%, which is arguably a relatively large value, but sufficient for our purpose. In order to perform the Monte-Carlo integration we smoothen the singularities of the delta-functions as well as the step-functions by replacing the delta-functions by Lorentzians with a FWHM of $0.01$ and the step functions by Fermi distributions at an effective inverse temperature $\beta=200$. 

The second step is to evaluate the real part of the self energy by a Kramers-Kronig transform and determine the dopon dispersion $\xi_\k$ by finding the maximum of the dopon spectral function
\begin{equation}
A(\k, \omega) = \frac{1}{\pi} \frac{-\text{Im} \Sigma_R(\k, \omega)}{[-\omega+\xi^{(0)}_\k+\text{Re} \Sigma_R(\k, \omega)]^2+[\text{Im} \Sigma_R(\k, \omega)]^2} \ .
\end{equation}
Here, $\xi^{(0)}$ denotes the Gaussian dopon dispersion from Equ.~\eqref{dopdisp}. The quasiparticle residue $Z_\k$ is obtained via
\begin{equation}
Z^{-1}_\k =\left| 1- \frac{\partial \, \text{Re}  \Sigma_R(\k,\omega)}{\partial \omega} \right|_{\omega=\xi_\k} \ .
\end{equation}
Finally, the self consistency loop is performed by plugging $\xi_\k$ and $Z_\k$ back into Equ.~\eqref{imsigma} and repeating the steps above until convergence is achieved.

\begin{figure}
\begin{center}
\includegraphics[width= .95 \textwidth]{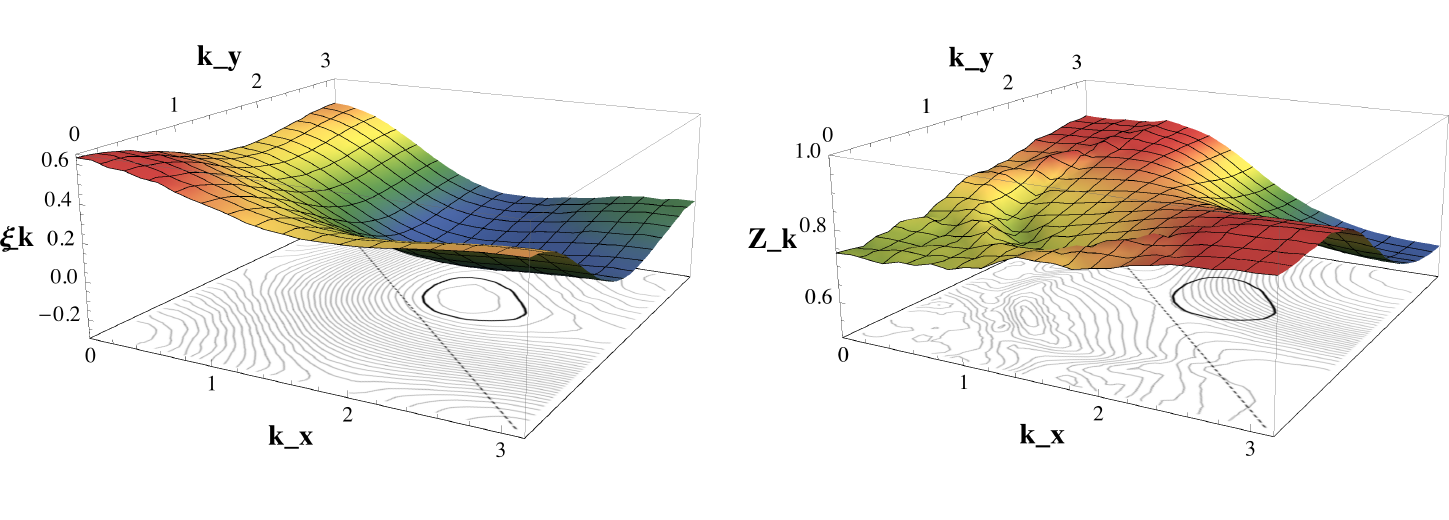}
\end{center}
\caption{(Color online) Self-consistent dopon dispersion $\xi_\k$ (left) and dopon quasiparticle residue $Z_\k$ (right) in the $\mathbb{Z}_2$-FL* phase as a function of $k_x$ and $k_y$ in the upper right quadrant of the Brillouin zone. The thick black contour marks the position of the dopon Fermi surface, which coincides with the electron Fermi surface. The dashed line indicates the magnetic Brillouin zone boundary. Parameter values for this plot are: $Q=0.4$, $\Delta=0.025$, $\mu=0.083$ and the rest as in Tab.~\ref{params}.}
\label{figdispZ}
\end{figure}

\subsection{Relation between the electron- and the dopon Fermi-surface}

Using Equ.~\eqref{ctilde} the electron momentum distribution can be expressed in terms of the lattice spin- and dopon operators as\begin{equation}
c^\dagger_{\k \sigma} c^{\ }_{\k \sigma} = \text{const.} + \frac{1}{2} \sum_{ij} e^{i \k \cdot (\R_i-\R_j)} \left[ -(1/4+\S_i\cdot \S_j ) d^\dagger_{j \sigma} d^{\ }_{i \sigma} + (d^\dagger_{j \alpha} \vec{\sigma}_{\alpha \beta} d^{\ }_{i \beta}) \cdot (\S_i + \S_j)/2 \right] \ ,
\end{equation}
where we implicitly sum over repeated spin indices and again neglect the $\S_i \times \S_j$ term.
Using the Schwinger-Boson representation \eqref{sb} for the lattice spins the electron momentum distribution is given by
\begin{eqnarray}
\langle c^\dagger_{\k \sigma} c^{\ }_{\k \sigma} \rangle &=& \text{const.} + \frac{1}{2} \sum_{ij} e^{i \k \cdot (\R_i-\R_j)} \Big\langle - d^\dagger_{j \sigma} d^{\ }_{i \sigma} + \frac{1}{2} (\epsilon_{\alpha \beta} b^\dagger_{i \alpha} b^{\dagger }_{j \beta}) (\epsilon_{\gamma \delta} b^{\ }_{i \gamma} b^{\ }_{j \delta}) d^\dagger_{j \sigma}  d^{\ }_{i \sigma} \notag \\
&& \hspace{4.75cm}+ \frac{1}{4} d^\dagger_{j \alpha}  d^{\ }_{i \beta} (b^\dagger_{i \beta} b^{\ }_{i \alpha} + b^\dagger_{j \beta} b^{\ }_{j \alpha})  \Big\rangle \notag \\
&=& -\frac{1}{2} \langle d^\dagger_{-\k \sigma} d^{\ }_{-\k \sigma} \rangle + \text{smooth function of} \ \k \ .
\label{momdistr}
\end{eqnarray}
The second line follows because the last two terms give rise to convolutions of the dopon momentum distribution with spinon correlators, where the Fermi surface singularity is smoothened out. We thus conclude that the electron Fermi surface coincides with the dopon Fermi surface. Moreover, the value of the electron quasiparticle residue at the Fermi surface is one half times the dopon quasiparticle residue $Z_\k$.

From Equ.~\eqref{momdistr} it is also clear that the spinon-dopon approach to the t-J model cannot give rise to electron pockets at the antinodal regions, as observed in recent experiments. Instead the Fermi surfaces are always hole-like.

\subsection{Results}

\begin{figure}
\begin{center}
\includegraphics[width= .95 \textwidth]{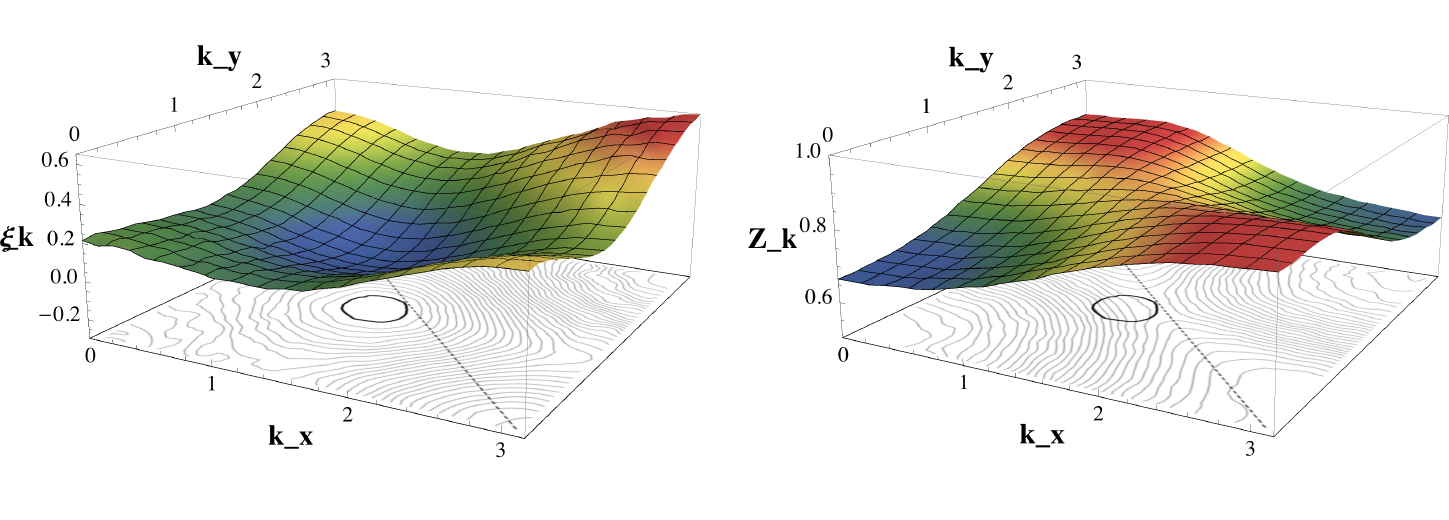}
\end{center}
\caption{(Color online) As in Fig.~\ref{figdispZ} but with parameters: $Q=0.54$, $\Delta=0.025$, $\mu=0.18$.}
\label{figdispZ2}
\end{figure}

The following results were obtained using standard values for the bare hopping amplitudes, shown in Tab.~\ref{params}. The nearest-neighbor hopping amplitude defines our energy scale and has been set to unity.  
\begin{table}[htdp]
\caption{parameter values}
\begin{center}
\begin{tabular}{c|c||c|c||c|c||c|c}
t & 1 & J & 0.25 & Q & variable & $\mu$ & variable \\
t' & -0.3 & J' & 0.05 & Q' & 0.1 & $\Delta$ & variable \\
t'' & 0.1 
\end{tabular}
\end{center}
\label{params}
\end{table}
The next nearest neighbor exchange interaction $J'$ as well as the corresponding singlet amplitude $Q'$ were chosen to be relatively small compared to the nearest neighbor values, as their only purpose is to break the IGG form $U(1)$ down to $\mathbb{Z}_2$.

\begin{figure}
\begin{center}
\includegraphics[width= .95 \textwidth]{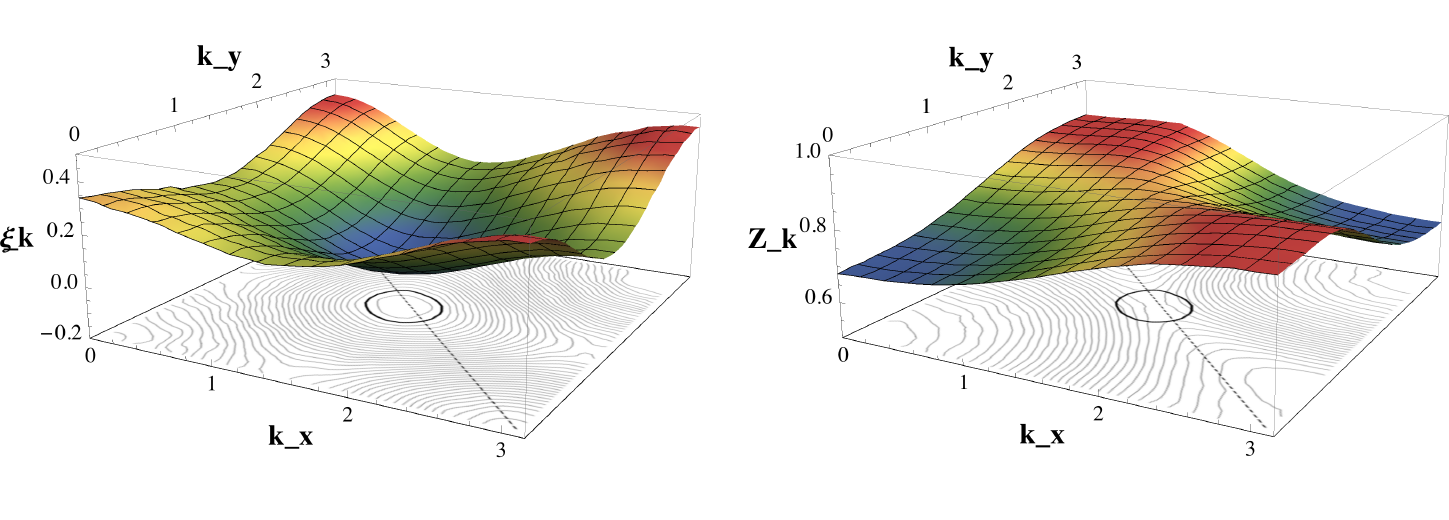}
\end{center}
\caption{(Color online) As in Fig.~\ref{figdispZ} but with parameters: $Q=0.5$, $\Delta=0.01$, $\mu=0.182$.}
\label{figdispZ3}
\end{figure}

\begin{figure}
\begin{center}
\includegraphics[width= .4 \textwidth]{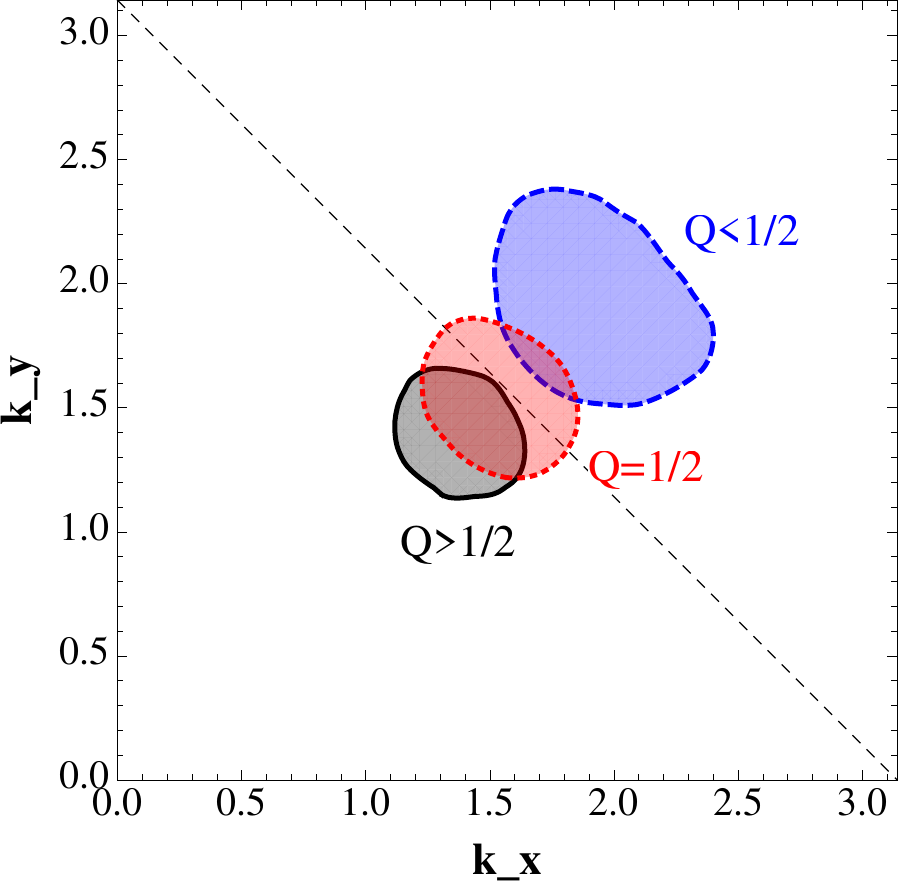}
\end{center}
\caption{(Color online) Evolution of the Fermi surface in the $\mathbb{Z}_2$-FL* phase as a function of the nearest neighbor singlet amplitude $Q\in[0, 1/\sqrt{2}]$. Shown is the upper right quadrant of the Brillouin zone. Black solid line: $Q=0.54, \Delta=0.025, \mu=0.18$, red dotted line: $Q=0.5, \Delta=0.01, \mu=0.182$, blue dashed line: $Q=0.4, \Delta=0.025, \mu=0.083$, other parameters as in Tab.~\ref{params}. The hole-pockets move to the inner side of the magnetic Brillouin zone boundary (indicated by the dashed line) as the strength of local antiferromagnetic correlations, parametrized by $Q$, increases.}
\label{figFS}
\end{figure}

Our results for the self-consistent dopon dispersion $\xi_\k$ and quasiparticle residue $Z_\k$ as a function of $k_x$ and $k_y$  in the upper right quadrant of the Brillouin zone are shown in Figs.~\ref{figdispZ}, \ref{figdispZ2} and \ref{figdispZ3}. These results are at a finite dopon density $n_d > 0$, although we note that $\xi_\k$ as well as $Z_\k$ look qualitatively similar in the case where the dopon density is going to zero. 
The position of the dispersion minimum depends on the strength of the local antiferromagnetic correlations, which is parametrized by the singlet bond-amplitude $Q$. For $Q=1/2$ the hole pockets are aligned with the magnetic Brillouin zone boundary and centered at $\q \simeq (\pi/2,\pi/2)$. For weaker correlations $Q < 0.5$ the hole-pockets are shifted to the outer side of the magnetic Brillouin zone boundary towards $\q=(\pi,\pi)$, whereas stronger local AF-correlations $Q>0.5$ give rise to hole-pockets centered on the inner side of the magnetic Brillouin zone boundary (see Fig.~\ref{figFS}). 

\begin{figure}
\begin{center}
\includegraphics[width= .4 \textwidth]{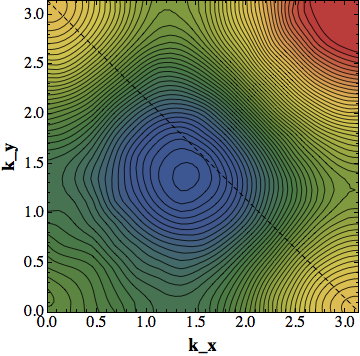}
\hspace{0.5cm}
\includegraphics[width= .4 \textwidth]{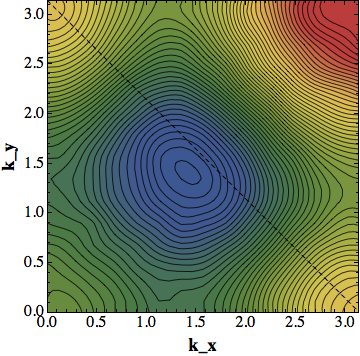}
\end{center}
\caption{(Color online) Evolution of the dopon dispersion $\xi_\k$ in the $\mathbb{Z}_2$-FL* phase as a function of the spinon gap $\Delta$ for $Q=0.54$ and $\mu=0.15$. Left: $\Delta=0.025$, right: $\Delta=0.005$, other parameters as in Tab.~\ref{params}. The Fermi energy is below the dopon band in both cases. With decreasing spinon gap $\Delta$ the dispersion around the minima becomes more elliptical.}
\label{figFS2}
\end{figure}

The spinon gap $\Delta$ does not influence the position of the pockets, but it changes their shape slightly. The smaller $\Delta$ is, the more elliptical are the hole-pockets. This is illustrated in Fig.~\ref{figFS2}. We note, however, that the ellipticity of the hole-pockets depends more strongly on the precise value of the bare hopping parameters $t'$ and $t''$ as on the size of the spinon gap. In fact, smaller $t'$ and $t''$ give rise to more elliptical pockets, similar to the standard SDW theory for antiferromagnetic metals.

The effective mass of the dopons turns out to be enhanced compared to the bare electron band mass as well. For the two dispersions shown in Fig.~\ref{figFS2}, the arithmetic mean of the effective masses at the dispersion minimum along the two principal axes is $\bar{m}_\text{eff} \approx 2.5$ in natural units (i.e.~$m=1$ corresponds to the band mass of the nearest neighbor tight-binding dispersion). Again the effective mass depends on the precise value of the bare hopping parameters $t'$ and $t''$. For highly elliptical pockets the effective mass can reach values on the order of $m_\text{eff} \sim 10$ along the flat direction.

Within our approximation scheme the quasiparticle residue $Z_\k$ does not drop sharply on the outer half of the Fermi surface, as expected from phenomenological models\cite{Qi}. Only for relatively high dopon fillings, as in Fig.~\ref{figdispZ}, an asymmetry of $Z_\k$ between the inner- and outer side of the hole-pocket appears. 

\section{Phase transition between a $\mathbb{Z}_2$-FL* and an AF-metal}
\label{sec:transition}

The Schwinger-Boson description \eqref{sb} is well suited to study the quantum phase transition between the $\mathbb{Z}_2$-FL* described above and a metal with long-range antiferromagnetic order. Indeed, AF-ordering corresponds to a condensation of Schwinger Bosons at the points where the spinon-gap closes. For $Q'=0$ the ordering wave vector is commensurate and the spinon dispersion \eqref{Ek} has two degenerate minima at the momenta $\q=\pm \K$ with $\K=(\pi/2, \pi/2)$. A condensate of the two Schwinger-Boson flavors at these respective momenta, i.e.~$\langle b_{\q \uparrow} \rangle = \sqrt{m_s} \, \delta_{\q,\K}$ and $\langle b_{\q \downarrow} \rangle = \sqrt{m_s} \, \delta_{\q,-\K}$, corresponds to an AF-ordered state with staggered magnetization $m_s$ in x-direction. 

\begin{figure}
\begin{center}
\includegraphics[width= .4 \textwidth]{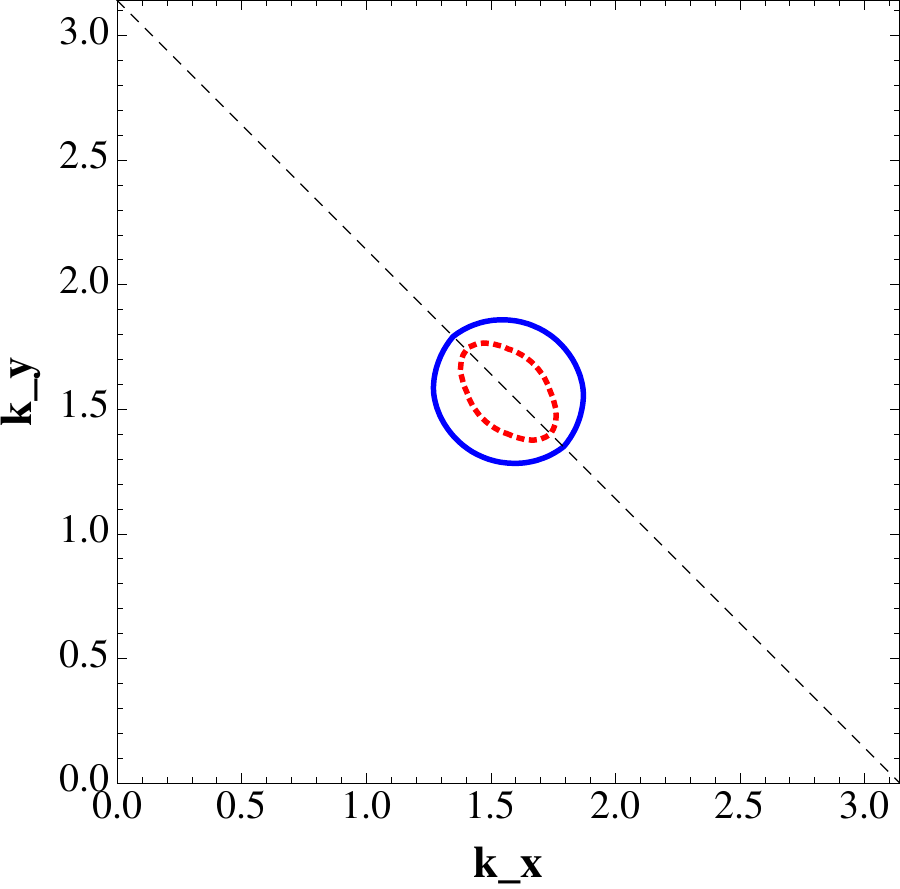}
\hspace{0.5cm}
\includegraphics[width= .4 \textwidth]{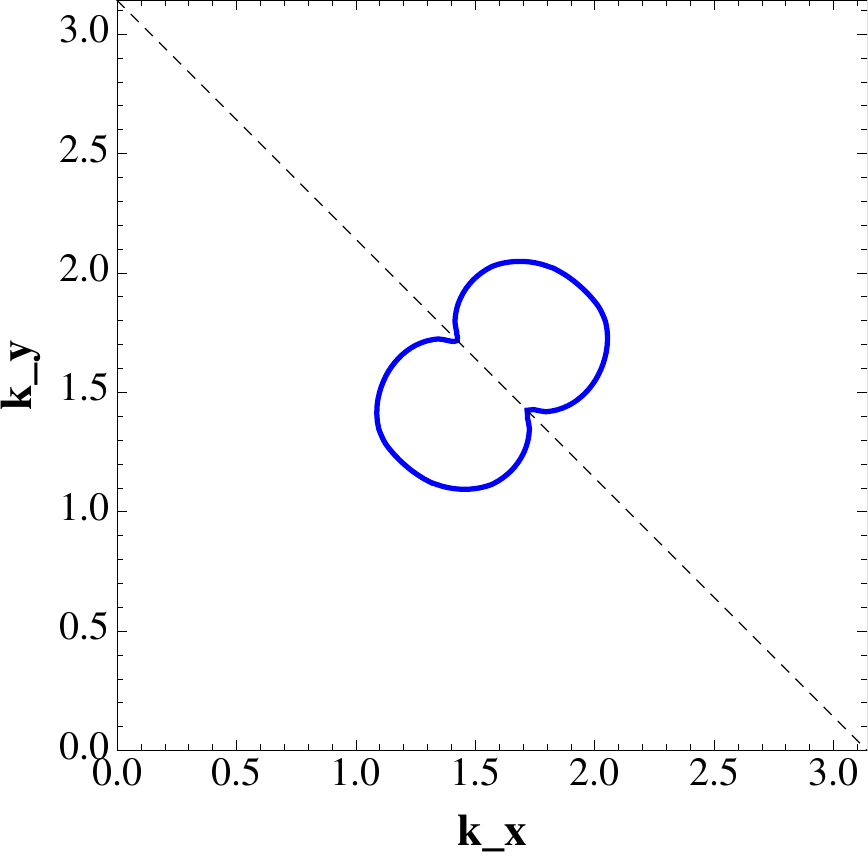}
\end{center}
\caption{(Color online) Possible Fermi surface shapes in the antiferromagnetically ordered phase for $m_s \ll 1$. Shown is the upper right quadrant of the Brillouin zone. The dashed line marks the magnetic Brillouin zone boundary. See text for a discussion.}
\label{figFSAF}
\end{figure}

Within our path integral formulation in Equ.~\eqref{hmf} such a Schwinger-Boson condensate can be straightforwardly introduced by shifting the Nambu fields 
\begin{equation}
B_{q \sigma} \to B_{q \sigma} + \sqrt{m_s} \,  \delta_{\Omega_n,0} \, \delta_{\q, \K} 
\end{equation}
and keeping only the quadratic terms in the shifted field. Note that the mean field ansatz $Q_\k$ in Equ.~\eqref{ansatz} also acquires a contribution from the condensate
\begin{equation}
Q_\k = \frac{m_s}{2} (\delta_{\k, \K} - \delta_{\k,-\K}) + Q^{(0)}_\k 
\label{ansatzAF}
\end{equation}
where $Q^{(0)}_\k = i 2 Q (\sin k_x + \sin k_y) $ describes the strong nearest-neighbor correlations on top of the uniform long-range correlations induced by the condensate.
The mean-field action in the AF-ordered phase has the same form as Equ.~\eqref{hmf} with three differences: first, $Q_\k$ is given by Equ.~\eqref{ansatzAF}. Second, the bare dopon dispersion $\xi^0_\k$ in Equ.~\eqref{xi0} is replaced by
\begin{equation}
\xi^0_\k = t_\k (1-m_s)-m_s^2 (t_\k-t_{\k-\boldsymbol{\pi}}) +2 \sum_{\q \p} {Q_\q^{(0)}}^* \, t_\p \, Q^{(0)}_{\k+\q-\p} - \mu 
\label{xi0AF}
\end{equation}
and most importantly, the condensate gives rise to an additional term to the action \eqref{hmf} which describes the scattering of dopons with momentum transfer $\q=\boldsymbol{\pi}$ and which takes the form
\begin{equation}
S_\text{AF}/\beta = - \frac{m_s}{2} \sum_{\omega_n,\k} (t_\k +t_{\k+\boldsymbol{\pi}}) (\bar{d}_{\omega_n, \k+\boldsymbol{\pi} \uparrow} \, d_{\omega_n, \k \downarrow} + \text{h.c.})
\end{equation}
Now we can perform the same analysis as in Sec.~\ref{sec:flstar} by integrating out the bosonic modes and performing a Hartree-Fock analysis of the effective quartic dopon action. At this level of approximation the off-diagonal elements of the self-energy in spin-space vanish identically and the diagonal elements have exactly the same structure as in Sec.~\ref{sec:flstar}. The effective dopon action in the AF-ordered phase including the self-energy corrections is thus given by
\begin{equation}
S/\beta = \sum_k \begin{pmatrix}
\bar{d}_{\omega_n, \k+\boldsymbol{\pi} \uparrow} & \bar{d}_{\omega_n, \k \downarrow} 
\end{pmatrix}
\begin{bmatrix}
\dfrac{-i \omega_n + \xi_{ \k + \boldsymbol{\pi} }}{Z_{ \k + \boldsymbol{\pi} }} & - \dfrac{m_s}{2} (t_\k + t_{\k+ \boldsymbol{\pi} })  \\
- \dfrac{m_s}{2} (t_\k + t_{\k+ \boldsymbol{\pi}}) & \dfrac{-i \omega_n +\xi_\k }{Z_\k}
\end{bmatrix}
\begin{pmatrix}
d_{\omega_n, \k+\boldsymbol{\pi} \uparrow} \\
d_{\omega_n, \k \downarrow} 
\end{pmatrix} \ ,
\label{actionAF}
\end{equation}
where $\xi_\k$ and $Z_\k$ again denote the self-consistently determined dopon dispersion and quasiparticle residue, calculated in the same manner as in Sec.~\ref{sec:flstar} using Eqs.~\eqref{ansatzAF} and \eqref{xi0AF}. Diagonalizing \eqref{actionAF} we obtain two dopon bands with dispersions
\begin{equation}
\omega_\pm(\k)= \frac{\xi_{ \k + \boldsymbol{\pi}}+\xi_{ \k}}{2} \pm \frac{1}{2} \sqrt{(\xi_{ \k + \boldsymbol{\pi}} - \xi_{ \k})^2 + m_s^2 Z_\k Z_{ \k + \boldsymbol{\pi}} \, (t_\k + t_{\k+ \boldsymbol{\pi} })^2} \ .
\end{equation}
The dopon Fermi surface is determined by $\omega_\pm(\k)= 0$. 
Slightly beyond the AF critical point, where the condensate density is small ($m_s \ll 1$), the self energy contributions to the dopon Green's function are basically the same as in the $\mathbb{Z}_2$-FL* phase with a vanishing spinon gap and the dispersion $\xi_\k$ has a minimum close to $\q=(\pi/2, \pi/2)$, as in Sec.~\ref{sec:flstar}. The dopon Fermi surface in the AF-ordered phase thus again takes the form of pockets close to $\q=(\pi/2, \pi/2)$. However, the pockets are symmetric with respect to the magnetic Brillouin zone (BZ) boundary due to the presence of long-range antiferromagnetic order. The shape of the hole-pocket again depends on the strength of the short-range correlations, parametrized by $Q$ in Equ. \eqref{ansatzAF}. Possible Fermi-surfaces for $m_s =0.05$ are shown in Fig.~\ref{figFSAF}. For these plots we used the same $\xi_\k$ and $Z_\k$ as in the FL* phase with a vanishingly small spinon gap and $Q'=0$, which is justified for $m_s \ll 1$, as argued above. Note that for $Q=1/2$ the dispersion $\xi_\k$ is almost symmetric with respect to the magnetic Brillouin zone boundary and thus the two dopon bands in the AF-ordered phase are almost degenerate for $m_s \ll 1$. In this case we get two concentric Fermi pockets, shown as solid blue and dashed red lines in the left plot of Fig.~\ref{figFSAF}. For larger $m_s$, the red Fermi pocket shrinks to zero, and we
eventually obtain the familiar single hole pocket centered on $\q=(\pi/2, \pi/2)$ of the AF ordered phase.
The right plot shows a Fermi surface for $Q=0.54$, in which case the single pocket on the inner side of the magnetic BZ in the FL* phase is ``symmetrized" at the magnetic BZ boundary. 

Note that in the AF-ordered phase the electron Fermi surface is related to the dopon Fermi surface not by the same Equ.~\eqref{momdistr} as in the FL* phase. The spinon condensate gives rise to additional contributions $\sim m_s$, which only change the electron quasiparticle residue, however. The shape of the electron Fermi surface still coincides with the dopon Fermi surface.

The nature of the quantum critical point between the AF-ordered and FL* phases can be addressed by methods
similar to earlier work\cite{css,morinari,ribhu,tarun}. The magnetic fluctuations are described by the spinor Schwinger bosons,
and their critical theory is the O(4) Wilson-Fisher fixed point. We now have to check if this critical point is destabilized by the dopon
Fermi surfaces. Because the dopons don't carry emergent gauge charges, they couple rather weakly to the critical spin 
fluctuations \cite{Qi}; the influence of this coupling can be analyzed perturbatively, and as in previous work \cite{morinari,tarun}
it is found to be irrelevant. So the critical theory remains that of the deconfined O(4) variety.\cite{css}

\section{Conclusions}
\label{sec:conc}

This paper has presented a microscopic construction of a FL* phase in a single-band $t$-$J$ model on the square lattice,
and described its evolution towards the onset of antiferromagnetic order. 
This was achieved by writing the $t$-$J$ model in a Kondo-like formulation using the spinon-dopon formalism.\cite{Ribeiro}
The FL* phase had a ``background'' spin liquid, which was the $\mathbb{Z}_2$ spin liquid with bosonic spinon excitations.
This spinon was then coupled to mobile carriers (the ``dopons'') which had the same quantum numbers of the electron.
Our effective Hamiltonian for the spinons and dopons was an exact, in principle, representation of the $t$-$J$ model.
However, we only analyzed this effective Hamiltonian in a relatively straightforward self-consistent one-loop approximation.
But there is an obstacle to extending such an analysis to higher orders and accuracy by using more powerful computational methods.

Despite its uncontrolled nature, our analysis yielded physically sensible results for the electron spectrum, which resemble aspects of the experimental observations. The key feature was the presence of a small hole pocket centered near but {\em not\/} at the magnetic Brillouin zone boundary. This pocket enclosed a volume determined by $x$, the density of doped carriers alone.
The quasiparticle residue was anisotropic around the Fermi surface, but our approximation did not yield the strong variation
found in earlier phenomenological models.\cite{Qi} 

We note a recent independent study\cite{weng} to describing the under-doped cuprates 
as a ``Luttinger-volume violating Fermi liquid'' (LvvFL) with a spin liquid of fermionic spinons.
The LvvFL state is qualitatively the same as the FL* state.

On the experimental front, there are a number of recent indications that FL*-like a model of pocket Fermi surfaces without
antiferromagnetic long-range order may be appropriate for the pseudogap region of the hole-doped cuprates.
The angle dependence of quantum oscillations
in YBa$_2$Cu$_3$O$_{6.59}$ is consistent \cite{ramshaw} with the absence of spin-density wave ordering.
NMR measurements \cite{julien} 
on YBa$_2$Cu$_3$O$_y$ have so far not seen antiferromagnetic order at fields as high as 30 Tesla,
but do report evidence of charge-ordering. Such a charge ordering can be superposed on
our FL* analysis in a straightforward manner; as long as the charge ordering wavevector does not connect the hole
pockets of the FL* state, there will be little change in the Fermi surface configuration.
And we have already noted photoemission evidence for pocket Fermi surfaces.\cite{Yang2011}

\acknowledgements

MP gratefully acknowledges numerous discussions with A.-M.~S.~Tremblay and P.~Strack.
This research was supported by the National Science Foundation under grant DMR-1103860, 
and by a MURI grant from AFOSR.
MP is supported by the Erwin Schr\"odinger Fellowship J 3077-N16 of the Austrian Science Fund (FWF).

\end{document}